\documentclass[aps,prc,preprint,groupedaddress,showpacs]{revtex4}

\usepackage{graphicx}
\usepackage{epstopdf}
\usepackage{amsmath}
\usepackage{multirow}
\usepackage{array}
\usepackage{subfigure}

\newcommand {\be}{\begin{equation}}
\newcommand {\ee}{\end{equation}}
\newcommand {\bea}{\begin{eqnarray}}
\newcommand {\eea}{\end{eqnarray}}

\newcommand{\btau}{\mbox{\boldmath$\tau$}}
\newcommand{\bpi}{\mbox{\boldmath$\pi$}}

\newcommand{\lqt}{\textquotedblleft}
\newcommand{\la}{\langle}
\newcommand{\ra}{\rangle}

\newcommand{\ddr}{\frac{\partial}{\partial r}}
\newcommand{\mo}{\mu(\omega)}

\begin{document}

\preprint{NT@UW-10-09}

\title{Impulse approximation in the $np\rightarrow d\pi^0$ reaction reexamined}

\author{Daniel R. Bolton}
\author{Gerald A. Miller}
\affiliation{Department of Physics, University of Washington, Seattle, Washington 98195-1560, USA}

\date{\today}

\begin{abstract}
The impulse approximation (one-body operator) in the $np\rightarrow d\pi^0$ reaction is reexamined with emphasis on the issues of reducibility and recoil corrections.  An inconsistency when one pion exchange is included in the production operator is demonstrated and then resolved via the introduction of \lqt wave function corrections" which nearly vanish for static nucleon propagators.  Inclusion of the recoil corrections to the nucleon propagators is found to change the magnitude and sign of the impulse production amplitude, worsening agreement with the experimental cross section by $\sim30\%$.  A cutoff is used to account for the phenomenological nature of the external wave functions, and is found to have a significant impact for $\Lambda\lesssim 2.5$ GeV.
\end{abstract}


\maketitle

\section{Introduction\label{sec:introduction}}

The theory of pion production in nucleon-nucleon collisions has seen many advances.  For an older review see Ref. \cite{Measday:1979if} and for a newer one see Ref. \cite{Hanhart:2003pg}.  Modern calculations are based on effective field theory (EFT) in an effort to eventually obtain model-independent predictions.  Such a goal is quite ambitious and naturally there are several major obstacles to overcome.  Two such obstacles are the large threshold momentum and the presence of initial and final state interactions.  Understanding pion production is interesting largely because of the aforementioned large threshold momentum.  When neutrons and protons with such momenta collide, they come much nearer to each other than they do in the similar $np\rightarrow d\gamma$ reaction.  Thus we are able to \lqt see" and therefore study a short-range region of strong nuclear force.

The EFT that has been used to describe $np\rightarrow d\pi^0$ is called baryon chiral perturbation theory \cite{Jenkins:1990jv, Weinberg:1991um, Bernard:1993nj, Hemmert:1997ye}, a theory based on the low-energy symmetries of QCD.  The theory consists of an infinite sum of interactions which are organized according to a \lqt power counting" scheme, an expansion in the typical momentum appearing in an interaction divided by the symmetry-breaking scale $\Lambda_\chi\approx m_N$.  This power counting was developed assuming the momentum to be $\sim m_\pi$.  In order to produce a pion at rest, the initial relative momentum of the colliding nucleons must be $p\sim\sqrt{m_\pi m_N}$ which means that the expansion parameter of the theory, $\chi\equiv\sqrt{m_\pi m_N}/m_N=0.38$, is much larger than the normal $m_\pi/m_N=0.14$.  It was first proposed in Ref. \cite{Cohen:1995cc} that the terms in the theory be reorganized to reflect this large momentum.  This modified power counting scheme is referred to as MCS.

In Ref. \cite{Lensky:2005jc}, MCS was used to answer two important questions: why the theoretical $pp\rightarrow d\pi^+$ (equivalent to $np\rightarrow d\pi^0$ under isospin symmetry) cross section was so much smaller than in experiment and why there existed formal inconsistencies with the next-to-leading-order (NLO) loops.  The answers to both questions followed from a subtle reducibility issue with the rescattering diagram that dominates the cross section.  In this paper we investigate the question of reducibility in the impulse approximation.

An attempt to address this issue was put forth in our recent study \cite{Bolton:2009rq} of charge symmetry breaking in $np\rightarrow d\pi^0$ where we introduced \lqt wave function corrections."  These corrections were calculated for the final state, and found to be a small fraction of the impulse diagram that they are correcting.  However, this calculation suffers from a particular approximation which we will describe.  Fixing this approximation has a very significant effect.  Furthermore, we show that the wave function corrections in the initial state are larger than one would expect in the MCS scheme.

In Sec. \ref{sec:reaction} we review the $np\rightarrow d\pi^0$ reaction and the impulse approximation's role.  Then, Sec. \ref{sec:ope} examines the inconsistency that is found when one includes static one pion exchange (OPE) with the impulse approximation.  Also in this section, wave function corrections are presented as a solution to the problem.  Section \ref{sec:recoil} discusses the correct implementation of the recoil corrections to the nucleon propagators.  The effects of including a cutoff are shown in Sec. \ref{sec:cutoff}.  Finally, we discuss the total cross section in Sec. \ref{sec:results} and conclude in Sec. \ref{sec:summary}.

\section{Pion production\label{sec:reaction}}

\begin{figure}
\centering
\includegraphics[height=2in]{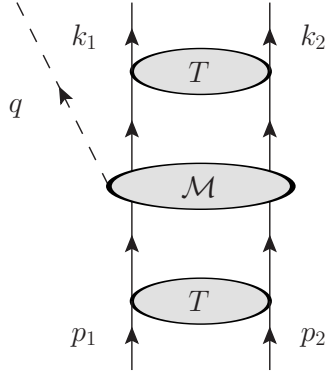}
\caption{Pion production operator.  Solid lines represent nucleons, dashed lines represent pions, and ovals represent interactions.\label{fig:momenta}}
\end{figure}

The pion production operator in momentum space depicted in Fig. \ref{fig:momenta} is a function of the pion momentum $\vec{q}$ and $\vec{l}=\vec{k}-\vec{p}$, where $\vec{k}$ ($\vec{p}$) refers to the final (initial) relative momentum of the nucleons.
The momentum transfer between the two nucleons in $\mathcal{M}$ is defined as $q'\equiv p_2-k_2$.  We use a phenomenological, non-relativistic potential ($V$) that is static: the energy of each individual nucleon is conserved by $V$.  Given this choice, working in the center of mass frame requires $q'\,^0=\omega_q/2$.  It should be noted that this is an approximation (the \lqt fixed kinematics approximation") that one needs to adopt in order to work in position space.  If one works in momentum space and fixes $q'\,^0$ via energy conservation at the $NN\pi$ vertex, this is called the \lqt equation of motion approximation."  As was shown in Ref. \cite{Hanhart:2000wf}, both of these approximations have problems, particularly when considering initial state interactions.  Nevertheless, it appears that the former is preferable to the latter.

\begin{figure}
\centering
\includegraphics[width=\textwidth]{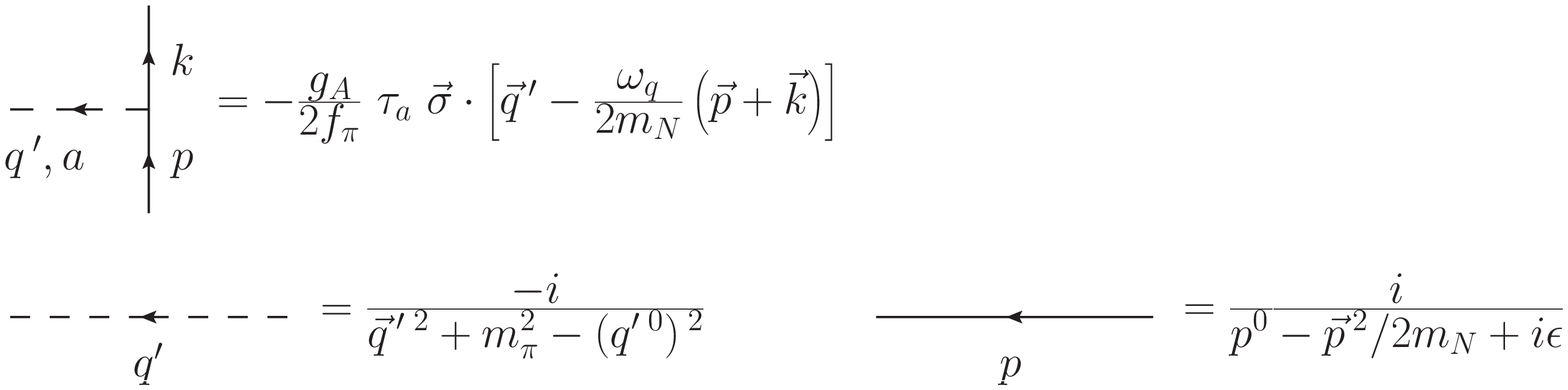}
\caption{Feynman rules\label{fig:rules}}
\end{figure}

In this work we employ threshold kinematics, where $q=(m_\pi,0)$, $q'=(m_\pi/2,\vec{l}\,)$, $p_{1,2}=(m_\pi/2,\pm\vec{p}\,)$, and $k_{1,2}=(0,\pm\vec{k}\,)$ with $|\vec{p}|=359\text{ MeV}$.  Also, at threshold only $s$-wave pions ($l_\pi=0$ with respect to the deuteron) are produced and the initial state is purely $^3P_1$.  We use the hybrid methodology introduced in Ref. \cite{Park:2000ct} where \lqt operators" (two-particle irreducible diagrams) are calculated perturbatively and then convolved with $NN$ wave functions which are obtained using phenomenological potentials.  Details of how this procedure is carried out for $NN\rightarrow NN\pi$ can be found in Sec. III of Ref. \cite{Bolton:2009rq}.  Appendix \ref{sec:lagrangian} discusses the chiral Lagrangian that defines the theory.  Using Eqs. (\ref{eq:l0}) and (\ref{eq:l1}), we obtain the Feynman rules shown in Fig. \ref{fig:rules}.

\begin{figure}
\includegraphics[height=2in]{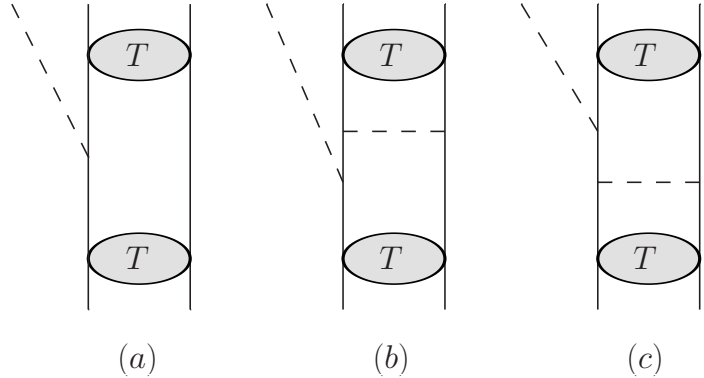}
\caption{Impulse approximation operator alone (a), with OPE pulled out from the final state (b), and with OPE pulled out from the initial state (c).  Solid lines represent nucleons, dashed lines represent pions, and ovals represent interactions.\label{fig:ia}}
\end{figure}

At leading order, $\mathcal{O}(\chi^1)$, the $s$-wave amplitude is dominated by the \lqt rescattering" diagram, where a single pion is emitted from one nucleon and inelastically scattered by the other nucleon into an on-shell-produced pion.  The only other leading order $s$-wave diagram is the impulse approximation, shown in Fig. \ref{fig:ia}(a) \cite{Bolton:2009rq}.
The details of the calculation of this diagram are given in Appendix \ref{sec:impdetails}.  For the initial and final states, we make use of three different potentials: Argonne v18 \cite{Wiringa:1994wb}, Nijmegen II \cite{Stoks:1994wp}, and Reid '93 \cite{Stoks:1994wp}.  The results for the $s$-wave reduced matrix element $A_0$ are shown in Table \ref{tab:iaresults} along with the leading order (LO) rescattering results which are detailed in Appendix C of Ref. \cite{Bolton:2009rq}.
\begin{table}
\caption{\label{tab:iaresults}Reduced matrix elements of the rescattering and impulse production operators for three different potentials.}
\begin{center}
\renewcommand{\tabcolsep}{3mm}
\begin{tabular}{|c||c|c|c|}
\hline
Diagram & Av18 & Nijm II & Reid '93\\ \noalign{\hrule height 1.5pt}
$A_0^{\text{res}}$ & 76.9 & 83.4 & 80.3\\ \hline
$A_0^{\text{imp}}$ & 4.9 & 1.3 & 3.5\\ \hline
\end{tabular}
\end{center}
\end{table}
$A_0$ also receives loop contributions at NLO ($\chi^2$) which were calculated in Ref. \cite{Hanhart:2002bu}.  However, a major result of the aformentioned Ref. \cite{Lensky:2005jc} was that these loops cancel, with the modification of the tree-level rescattering diagram being put on-shell.  This modification (which is truly LO) is included in Table \ref{tab:iaresults}.  Finally, we point out that the experimental data (Sec. \ref{sec:results}) imply a reduced matrix element of $80\leq A_0\leq94$.

It is important to notice that the impulse approximation operator of Fig. \ref{fig:ia}(a) cannot formally be convolved with the initial and final states as described above because the nucleon emitting the pion cannot remain on-shell.  To put it another way, $q'\,^0=0$ because there is no way for the energy to be transferred.  The common approximation made in pion production calculations is to ignore this formal difficulty.  On the other hand, one \lqt pulls" an OPE from the final state wave function [Fig. \ref{fig:ia}(b)] in order to argue that the impulse approximation is leading order.  This can be seen as an application of the Lippmann-Schwinger equation for the final state deuteron,
\be
\mid\psi_d\ra=GV\mid\psi_d\ra,\label{eq:LS}
\ee
where $G$ represents the two-nucleon propagator and $V$ represents the full potential.  In Figs. \ref{fig:ia}(b) and \ref{fig:ia}(c), we have made the approximation that $V\approx\text{OPE}$, which is known to be a good approximation for the deuteron \cite{Ericson:1988gk}, but is not as valid for the initial state.  Eq. (\ref{eq:LS}) begs the question: are the diagrams in Figs. \ref{fig:ia}(a) and \ref{fig:ia}(b) the same size?  If so, we will be able to conclude that Fig. \ref{fig:ia}(a) is doing its job well.

\section{Including OPE\label{sec:ope}}

Calculation of Fig. \ref{fig:ia}(b) using Eq. (\ref{eq:LS}) is detailed in Appendix \ref{sec:opedetails}.  For lack of a better name, we will call this the \lqt OPE reducible" diagram.  In this calculation, we take $G=(E-H_0)^{-1}=(-E_d-\vec{p}\,^2/m_N)^{-1}$ where $E_d=2.22$ MeV is the binding energy of the deuteron and $m_N/2$ is the reduced mass.  The energy of the exchanged pion in this case is taken to be $q'\,^0=0$.  This choice is consistent with the fact that the OPE is the first term in the $V$ of Eq. (\ref{eq:LS}), which should be the same $V$ that is used to generate the initial and final wave functions.  The results are shown in the second row of Table \ref{tab:opef}.
\begin{table}
\caption{\label{tab:opef}Reduced matrix elements of the impulse approximation with an OPE pulled out from the \textit{final} state interaction [see Fig. \ref{fig:ia}(b)].}
\begin{center}
\renewcommand{\tabcolsep}{3mm}
\begin{tabular}{|c||c|c|c|}
\hline
Diagram & Av18 & Nijm II & Reid '93\\ \noalign{\hrule height 1.5pt}
$A_0^{\text{OPE,red,f}}$ & 75.2 & 64.6 & 79.3\\ \hline
$A_0^{\text{OPE,irr,f}}$ & 75.6 & 64.7 & 79.8\\ \hline
$A_0^{\text{OPE,irr,f}}-A_0^{\text{OPE,red,f}}$ & 0.5 & 0.1 & 0.5\\ \hline
\end{tabular}
\end{center}
\end{table}

We find an inconsistency between the impulse approximation [Fig. \ref{fig:ia}(a)] and OPE reducible [Fig. \ref{fig:ia}(b)] diagrams: although they are equivalent according to the Lippmann-Schwinger equation, they are of very different size numerically.  Using Av18, they are 4.9 and 75.2, respectively.  Of course this inconsistency is not surprising when one notes that three-momentum transfer is provided for in the latter diagram but not the former.

To resolve this problem, we re-consider the diagram in Fig. \ref{fig:ia}(b) as a fully irreducible operator (without mentioning external wave functions).  This is justifiable in that the left intermediate nucleon is off-shell by $m_\pi$, more than the $m_\pi^2/m_N$ typical of reducible diagrams.  If we view it in this way, we are free to chose the energy of the exchanged pion to be $q'\,^0=m_\pi/2$ as mentioned in Sec. \ref{sec:reaction}.  Additionally, the single-nucleon propagator for the left intermediate nucleon is taken from the rules shown in Fig. \ref{fig:rules}.  The reduced matrix element for this \lqt OPE irreducible" operator is given in Appendix \ref{sec:opedetails}, and the results are shown in the third row of Table \ref{tab:opef}.  We find that this diagram, which correctly accounts for energy transfer, is approximately equal to the OPE reducible diagram.

The question remains: should Fig. \ref{fig:ia}(b) be included, and if so, how?  Until a clear procedure is defined for going from the full four-dimensional $\pi NN$ coupled-channels formalism to the more common three-dimensional uncoupled formalism, this question is open to interpretation.  We continue to take the view proposed in Ref. \cite{Bolton:2009rq}, which is that the OPE irreducible diagram should be included with the OPE reducible diagram subtracted off to prevent double counting.  This difference is shown in the fourth row of Table \ref{tab:opef}, and is referred to as the (nearly vanishing) \lqt wave function correction."  Note that the cancellation is not as trivial as it appears.  Schematically, OPE is two derivatives on a Yukawa that has different ranges for the reducible and irreducible cases.  In going from reducible to irreducible, the radial integral gets bigger because the range increases.  The derivatives bring down inverse powers of the range such that the overall amplitudes are similar in size.

As in Sec. \ref{sec:reaction}, for $s$-wave pions at threshold we have a $\vec{\sigma}\cdot(\vec{p}_i+\vec{p}_f)$ at the vertex where the pion is produced.  For this reason the authors of Ref. \cite{Bolton:2009rq} only considered OPE in the final state, assuming initial state OPE [Fig. \ref{fig:ia}(c)] to be suppressed by the small final state momentum $\vec{k}$.  However, one needs to be careful when applying power counting to calculations that involve external $NN$ wave functions.  At small distances the momenta of the nucleons (derivatives in position space) are distorted away from their constant values at asymptotically large distances.  As an example of this difficulty, it can be shown that the $\vec{k}\,^2/2m_N$ operator becomes larger than the $\vec{p}\,^2/2m_N$ operator in the rescattering diagram.  For this reason, we also calculate Fig. \ref{fig:ia}(c) (for the details, see Appendix \ref{sec:opedetails}).  The results of this calculation are shown in Table \ref{tab:opei} where, again, the full wave function correction is nearly zero.
\begin{table}
\caption{\label{tab:opei}Reduced matrix elements of the impulse approximation with an OPE pulled out from the \textit{initial} state interaction [see Fig. \ref{fig:ia}(c)].}
\begin{center}
\renewcommand{\tabcolsep}{3mm}
\begin{tabular}{|c||c|c|c|}
\hline
Diagram & Av18 & Nijm II & Reid '93\\ \noalign{\hrule height 1.5pt}
$A_0^{\text{OPE,red,i}}$ & -11.2 & -23.7 & -15.0\\ \hline
$A_0^{\text{OPE,irr,i}}$ & -11.2 & -23.7 & -15.0\\ \hline
$A_0^{\text{OPE,irr,i}}-A_0^{\text{OPE,red,i}}$ & $\sim$0 & $\sim$0 & $\sim$0\\ \hline
\end{tabular}
\end{center}
\end{table}

\begin{figure}
\centering
\includegraphics[height=1.5in]{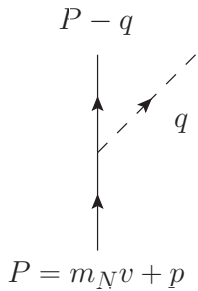}
\caption{Momenta of the nucleon propagator where $v=(1,0,0,0)$ is used\label{fig:nprop}}
\end{figure}

There is one more formal point to discuss with regard to the above calculations.  Expressions for nucleon propagators in irreducible diagrams differ based on the power counting scheme used.  Consider the situation shown in Fig. \ref{fig:nprop}.
Starting from the full, relativistic propagator, the authors of Ref. \cite{Hanhart:2007mu} showed that, if $p^0\sim m_\pi$ and $\vec{p}\sim\sqrt{m_\pi m_N}$, the correct propagator after emitting the pion is
\be
\frac{i}{-q^0+(2\vec{p}\cdot\vec{q}-\vec{q}\,^2)/2m_N+i\epsilon},\label{eq:mcsprop}
\ee
where the second term in the denominator is the recoil correction, suppressed by one power of $\chi$ if $q^0\sim\sqrt{m_\pi m_N}$.  Note that we have ignored the resulting vertex corrections and antinucleon effects, both of which are suppressed for any choice of $q^0$.  Equation (\ref{eq:mcsprop}), which we will refer to as the \lqt new" method, comes in opposition to the \lqt old" method which derives the propagator from the non-relativistic chiral Lagrangian,
\be
\frac{i}{p^0-q^0-(\vec{p}-\vec{q})^2/2m_N+i\epsilon},\label{eq:wprop}
\ee
where the third term comes from the NLO Lagrangian and is a candidate for promotion in MCS counting.  Let us now consider the case of Fig. \ref{fig:ia}(b), where in terms of the momenta shown in Fig. \ref{fig:nprop} we have $p=(m_\pi/2,\vec{p})$ and $q=(m_\pi,0)$.  Since $p^0-q^0\sim m_\pi$, we promote the recoil corrections in the old propagator and find
\bea
iG_\text{new}^\text{irr,f}&=&\frac{i}{-m_\pi}\nonumber
\\
iG_\text{old}^\text{irr,f}&=&\frac{i}{m_\pi/2-m_\pi-\vec{p}\,^2/2m_N}.
\eea
Next, consider the case of Fig. \ref{fig:ia}(c) where we have $p=(m_\pi/2,\vec{p})$ and $q=(-\omega,\vec{p}-\vec{k})$, and thus
\bea
iG_\text{new}^\text{irr,i}&=&\frac{i}{\omega+(\vec{p}\,^2-\vec{k}\,^2)/2m_N+i\epsilon}\nonumber
\\
iG_\text{old}^\text{irr,i}&=&\frac{i}{m_\pi/2+\omega-\vec{k}\,^2/2m_N+i\epsilon}.\label{eq:newoldi}
\eea
Note that in the absence of distortions ($|\vec{p}|\approx\sqrt{m_\pi m_N}$, $|\vec{k}|\sim0$), $iG_\text{new}=iG_\text{old}$ for both the initial and final state propagators.  For the sake of clarity we will define as the \lqt free recoil approximation" (FRA) the use of these free particle values for the nucleon momenta.

\section{Nucleon propagator recoil\label{sec:recoil}}

In Sec. \ref{sec:ope}, the FRA was used for the recoil corrections to the nucleon propagators.  In this section we calculate the diagrams again, treating the momenta properly as operators instead of numbers.  In this section we use the old nucleon propagators for the irreducible diagrams.  In doing so, we avoid the $G^i_\text{new}$ of Eq. (\ref{eq:newoldi}), which would be difficult to evaluate exactly in position space.  It was pointed out in Ref. \cite{Bernard:1998sz} that the old nucleon propagators have formal convergence problems owing to the large external momenta.  Nevertheless, we expect to gain insight into the validity of the FRA using these propagators,
\bea
iG^\text{irr,f}&=&\frac{i}{-m_\pi/2-\vec{p}\,^2/2m_N},\nonumber
\\
iG^\text{irr,i}&=&\frac{i}{m_\pi-\vec{k}\,^2/2m_N+i\epsilon}.\label{eq:npropnew}
\eea
Of course, according to MCS, $\vec{k}$ should not be counted as $\sim\sqrt{m_\pi m_N}$ and the $\vec{k}\,^2/2m_N$ term should therefore not appear as in Eq. (\ref{eq:npropnew}) until higher order.  We choose to retain it here as an investigation into the effects of the distortions.  For the reducible diagrams we continue to use $G^\text{red}=(E-H_0)^{-1}$.

The matrix elements can be calculated exactly in position space with Green function methods (see Appendix \ref{sec:recoildetails}).  The results are shown in Table \ref{tab:recoilprop}.
\begin{table}
\caption{\label{tab:recoilprop}Reduced matrix elements for the wave function corrections with proper treatment of the momenta using the old expressions for the nucleon propagators.}
\begin{center}
\renewcommand{\tabcolsep}{3mm}
\begin{tabular}{|c||c|c|c|}
\hline
& Av18 & Nijm II & Reid '93\\ \noalign{\hrule height 1.5pt}
$A_0^{\text{OPE,irr,f}}$ & 80.8 & 70.8 & 89.1\\ \hline
$A_0^{\text{OPE,red,f}}$ & 92.4 & 81.2 & 103.3\\ \hline
$A_0^{\text{OPE,irr,f}}-A_0^{\text{OPE,red,f}}$ & -11.7 & -10.4 & -14.2\\ \noalign{\hrule height 1.5pt}
$A_0^{\text{OPE,irr,i}}$ & 5.1+24.5$i$ & 15.1+34.7$i$ & 7.8+27.8$i$ \\ \hline
$A_0^{\text{OPE,red,i}}$ & 16.2+8.2$i$ & 22.9+9.9$i$ & 18.2+8.7$i$ \\ \hline
$A_0^{\text{OPE,irr,i}}-A_0^{\text{OPE,red,i}}$ & -11.1+16.3$i$ & -7.7+24.8$i$ & -10.4+19.1$i$ \\ \hline
\end{tabular}
\end{center}
\end{table}
We find that the final state wave function correction evaluated without the FRA gives a sizable negative contribution of approximately $-10$.  Additionally, we find that the initial state corrections become as important as the \textit{lower-order} final state corrections.  To verify the surprising results of this calculation, we examine as an example the $m_N\rightarrow\infty$ limit of the radial integral for the irreducible initial state OPE in comparison with its analog from the previous section (which is independent of $m_N$ since $\vec{k}=0$ is used),
\be
\mathcal{I}\equiv\int dr\,r^2\left(\sqrt{2}\ddr\frac{u(r)}{r}+\left(\ddr+\frac{3}{r}\right)\frac{w(r)}{r}\right)G^\text{OPE,irr,i}\left(2f(m_\pi/2,r)+g(m_\pi/2,r)\right)R_i(r),
\ee
where the functions $f$ and $g$ are defined in Appendix \ref{sec:opedetails}.  As shown in Fig. \ref{fig:largeM}, in the large-$m_N$ limit the recoil term in the propagator vanishes and we recover the leading order result.
\begin{figure}
\centering
\includegraphics[height=1.5in]{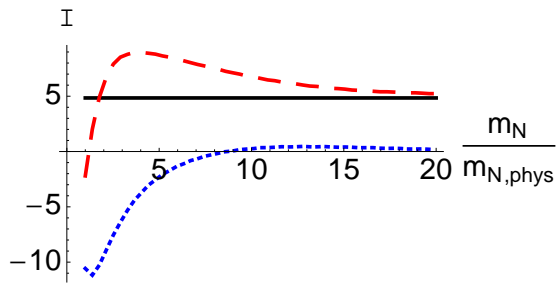}
\caption{\label{fig:largeM}(Color online) Irreducible initial state OPE integral as a function of $m_N$ using Av18.  The solid line displays the FRA result and the real (imaginary) part of the exact propagator result is shown as a dashed (dotted) curve.}
\end{figure}

\section{\label{sec:cutoff}Cutoff dependence}

It should not come as a surprise that odd things are happening in the short-distance part of the wave function, especially when we take derivatives.  In the hybrid formalism we are using, this domain of the wave function is calculated from a phenomenological potential: Woods-Saxon for Av18, one boson exchange for Nijm II, and Yukawa for Reid '93 (the very short range is exponential, exponential, and dipole, respectively).  Because these potentials are fitted to experimental phase shifts, the wave functions derived from them can be considered as infinitely high order in the EFT.  Thus one should consider using a cutoff to account for this mismatch between the operator and the wave functions.  Use of such a cutoff is referred to as EFT*, and was introduced in Ref. \cite{Park:2000ct}.

\begin{figure}
\centering
\includegraphics[width=.4\linewidth]{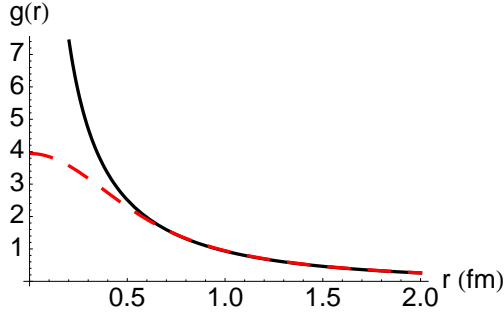}
\caption{\label{fig:cutoffyukawa}(Color online) Comparison of $g$ (solid curve) and $g_\Lambda$ (dashed curve) with $\omega=m_\pi/2$ and $\Lambda=1\text{ GeV}$.}
\end{figure}

\begin{figure}
\begin{minipage}{.4\linewidth}
\includegraphics[width=\linewidth]{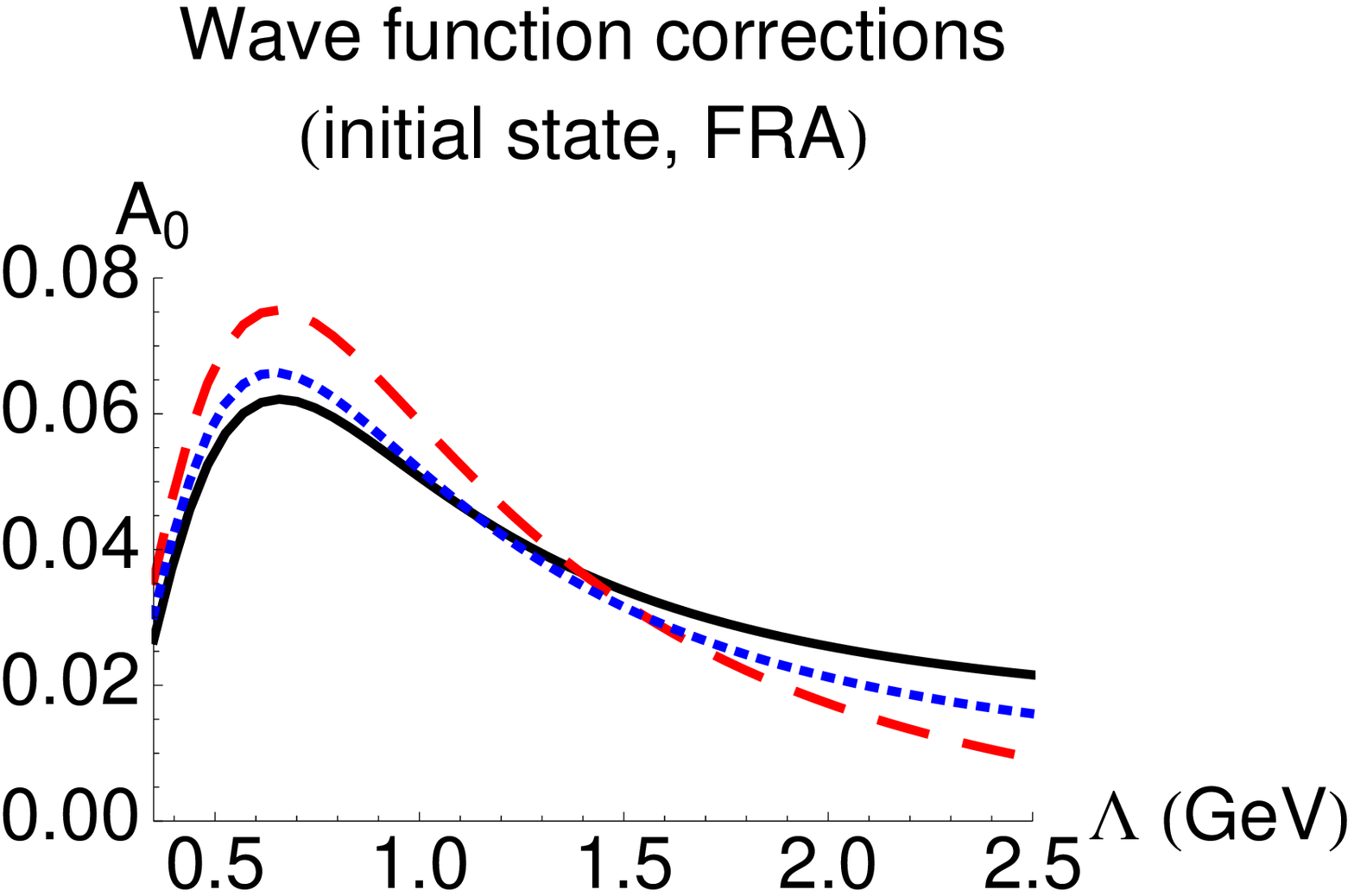}
\vspace{.05in}
\end{minipage}
\hspace{.1\linewidth}
\begin{minipage}{.4\linewidth}
\includegraphics[width=\linewidth]{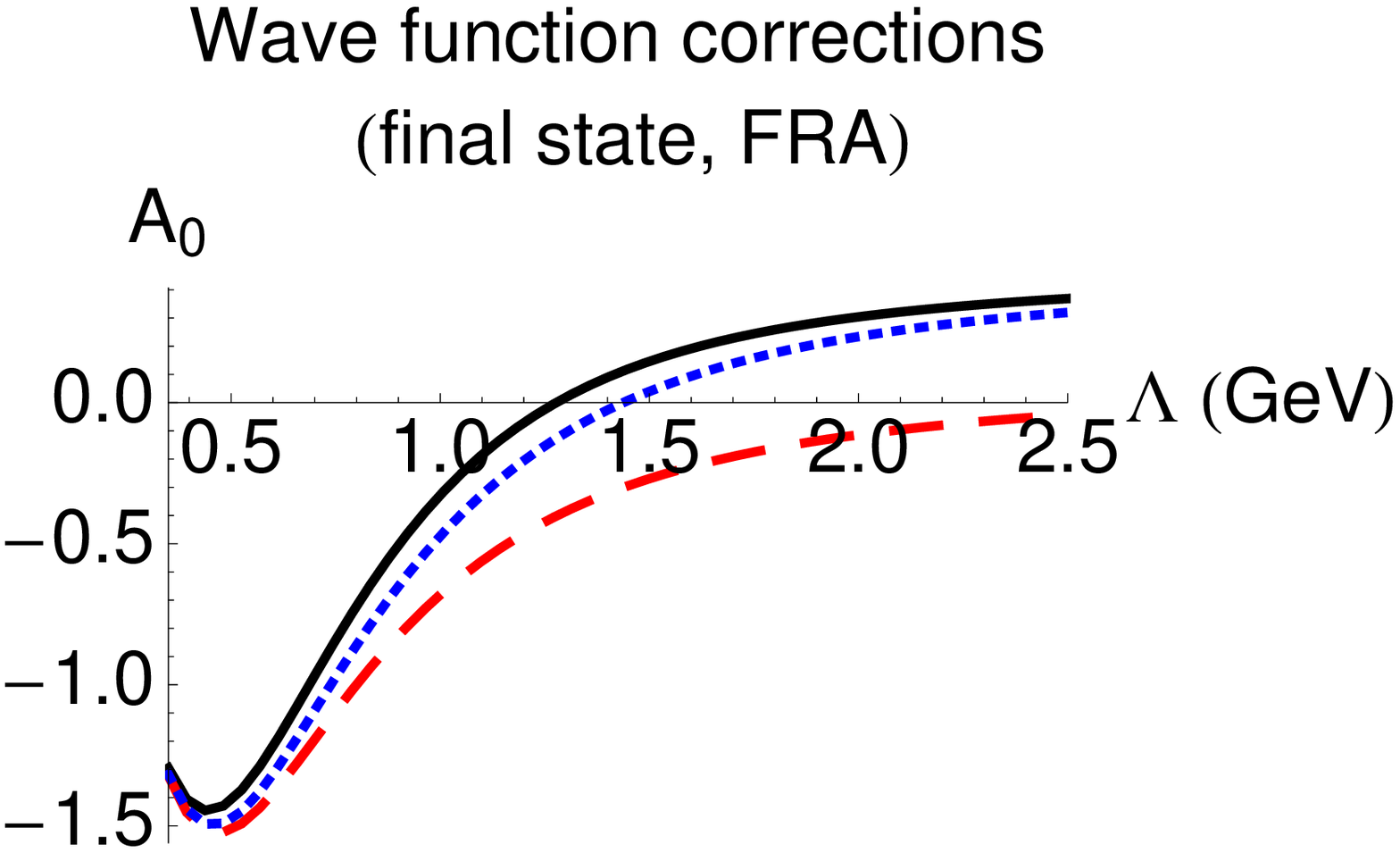}
\vspace{.05in}
\end{minipage}
\begin{minipage}{.4\linewidth}
\includegraphics[width=\linewidth]{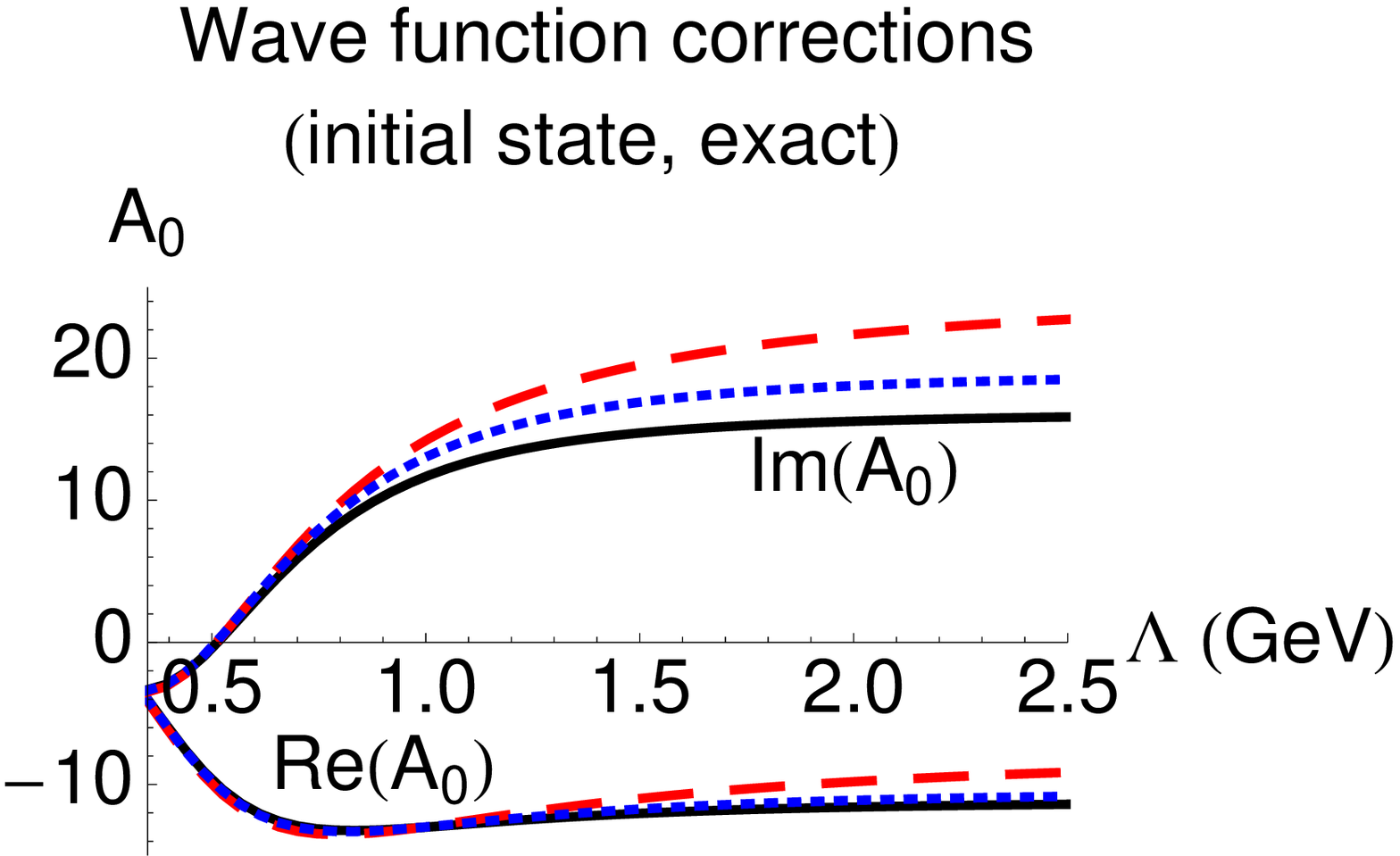}
\vspace{.05in}
\end{minipage}
\hspace{.1\linewidth}
\begin{minipage}{.4\linewidth}
\includegraphics[width=\linewidth]{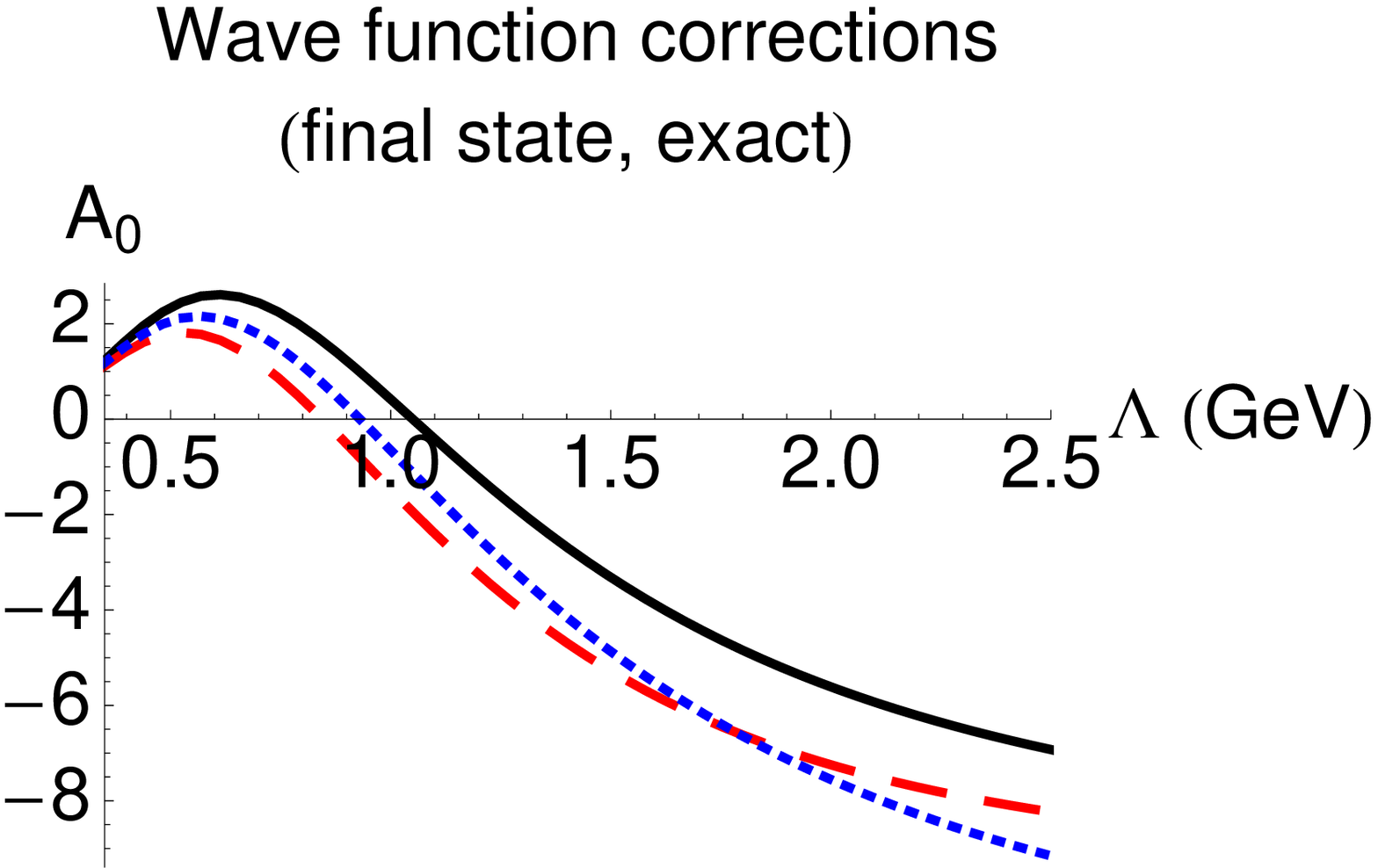}
\vspace{.05in}
\end{minipage}
\caption{\label{fig:cutoffmxels}(Color online) Cutoff dependence of various reduced matrix elements for Av18 (solid curve), NijmII (dashed curve), and Reid '93 (dotted curve).}
\end{figure}

In this section we investigate the effects of cutting off the convolution integrals that account for the the presence of initial and final states as discussed in Sec. \ref{sec:recoil}.  We use the procedure of Ref. \cite{Park:2002yp}, which modifies the Fourier transforms with a Gaussian cutoff,
\bea
\mathcal{M}(\vec{r})&=&\int\frac{d^3l}{(2\pi)^3}e^{i\vec{l}\cdot\vec{r}}S_\Lambda^2\left(\vec{l}\,^2\right)\mathcal{M}(\vec{l})\nonumber
\\
S_\Lambda\left(\vec{l}\,^2\right)&=&\exp\left(-\frac{\vec{l}\,^2}{2\Lambda^2}\right).
\eea
Note that the impulse approximation is not affected by such a cutoff scheme.  For the OPE operators we define $g_\Lambda(\omega,r)$:
\be
\frac{\mo g_\Lambda(\omega,r)}{4\pi}\equiv\int\frac{d^3l}{(2\pi)^3}e^{i\vec{l}\cdot\vec{r}-\vec{l}\,^2/\Lambda^2}\frac{1}{\vec{l}\,^2+\mo^2}.\label{eq:gLdef}
\ee
The exact evaluation of this integral and of the derivatives required to compute the diagrams of this work are shown in Appendix \ref{sec:cutoffdetails}.  As desired, the cutoff regulates the behavior of $g(r)$ at the origin, as shown in Fig. \ref{fig:cutoffyukawa}.
The cutoff dependence of various reduced matrix elements is shown in Fig. \ref{fig:cutoffmxels}.

The fact that we observe significant cutoff dependence of the wave function corrections above the typical scale of $\sim$1 GeV is surprising.  Indeed, this sensitivity indicates the need for a counterterm because observables must be cutoff independent.  As pointed out in Ref. \cite{Gardestig:2005sn}, if one considers the difference of terms that comprise the wave function correction, 
\be
\frac{1}{\vec{q}\,'\,^2+m_\pi^2}-\frac{1}{\vec{q}\,'\,^2+3m_\pi^2/4}=-\frac{m_\pi^2/4}{\vec{q}\,'\,^2+m_\pi^2}\cdot\frac{1}{\vec{q}\,'\,^2+3m_\pi^2/4},
\ee
it can be argued that wave function correction is N$^2$LO in the MCS scheme.  However, if this view is to be accepted, the fact that the wave function corrections are much larger in magnitude than the LO impulse approximation should be considered surprising.

\section{\label{sec:results}Cross section results}

Shown in Table \ref{tab:results} is a summary of the findings discussed in this paper at $\Lambda=\infty$ and $\Lambda=1$ GeV along with the rescattering diagram.
\begin{table}
\caption{\label{tab:results}Reduced matrix elements for three different potentials.}
\begin{center}
\renewcommand{\tabcolsep}{3mm}
\resizebox{\textwidth}{!}{
\begin{tabular}{|c||c|c|c|c|c|c|}
\hline
\multirow{2}{*}{Diagram} & \multicolumn{2}{c|}{Av18} & \multicolumn{2}{c|}{Nijm II} & \multicolumn{2}{c|}{Reid '93}\\
& $\Lambda=\infty$ & $\Lambda=1\text{ GeV}$ & $\Lambda=\infty$ & $\Lambda=1\text{ GeV}$& $\Lambda=\infty$ & $\Lambda=1\text{ GeV}$\\ \noalign{\hrule height 1.5pt}
Rescattering (NLO) & 76.9 & 76.2 & 83.4 & 81.5 & 80.3 & 79.1\\ \hline
Impulse & 4.9 & 4.9 & 1.3 & 1.3 & 3.5 & 3.5\\ \hline
Final wfn cor (FRA) & 0.5 & -0.3 & 0.1 & -0.7 & 0.5 & -0.5\\ \hline
Final wfn cor (exact) & -11.7 & 0.4 & -10.4 & -1.5 & -14.2 & -0.7\\ \hline
Initial wfn cor (FRA) & $\approx0$ & 0.1 & $\approx0$ & 0.1 & $\approx0$ & 0.1\\ \hline
Initial wfn cor (exact) & -11.1+16.3$i$ & -13.0+11.7$i$ & -7.7+24.8$i$ & -13.0+14.2$i$ & -10.4+19.1$i$ & -13.0+13.0$i$\\ \noalign{\hrule height 1.5pt}
Total (FRA) & 82.2 & 80.8 & 84.8 & 82.2 & 84.2 & 82.2\\ \hline
Total (exact) & 59.0+16.3$i$ & 68.5+11.7$i$ & 66.7+24.8$i$ & 68.4+14.2$i$ & 59.2+19.1$i$ & 69.0+13.0$i$\\ \hline
\end{tabular}
}
\end{center}
\end{table}
We also discuss the total cross section, which near threshold is parametrized as
\be
\sigma=\frac{1}{2}\left(\alpha\eta+\beta\eta^3\right),
\ee
where $q=m_\pi\eta$.  At threshold, one can only calculate $\alpha$,
\be
\alpha=\frac{m_\pi}{128\pi^2sp}|A_0|^2,\label{eq:alpha}
\ee
where $s=(m_\pi+m_d)^2$.  Note that charged pion production is related to neutral pion production by isospin symmetry (breaking is expected to be small in the total cross section).  This symmetry is the reason for the $1/2$ present in the definition of $\sigma$.  The most recent experimental data are shown in Table \ref{tab:expt}.
\begin{table}
\caption{\label{tab:expt}Experimental total cross section parameters}
\begin{center}
\renewcommand{\tabcolsep}{3mm}
\begin{tabular}{|c||c|c|}
\hline
Experiment & $\alpha\ (\mu\text{b})$ & $\beta\ (\mu\text{b})$\\ \noalign{\hrule height 1.5pt}
$np\rightarrow d\pi^0$ \cite{Hutcheon:1989bt} & $184\pm5$ & $781\pm79$\\ \hline
$\vec{p}p\rightarrow d\pi^+$ (Coulomb corrected) \cite{Heimberg:1996be} & $208\pm5$ & $1220\pm100$\\ \hline
$pp\rightarrow d\pi^+$ (Coulomb corrected) \cite{Drochner:1998ja} & $205\pm9$ & $791\pm79$\\ \hline
Pionic deuterium decay \cite{Strauch:2010rm} & $252^{+5}_{-11}$ & N/A\\ \hline
\end{tabular}
\end{center}
\end{table}

\begin{figure}
\begin{center}
\subfigure[\ FRA nucleon propagators]{\label{fig:totL-a}\includegraphics[width=.4\linewidth]{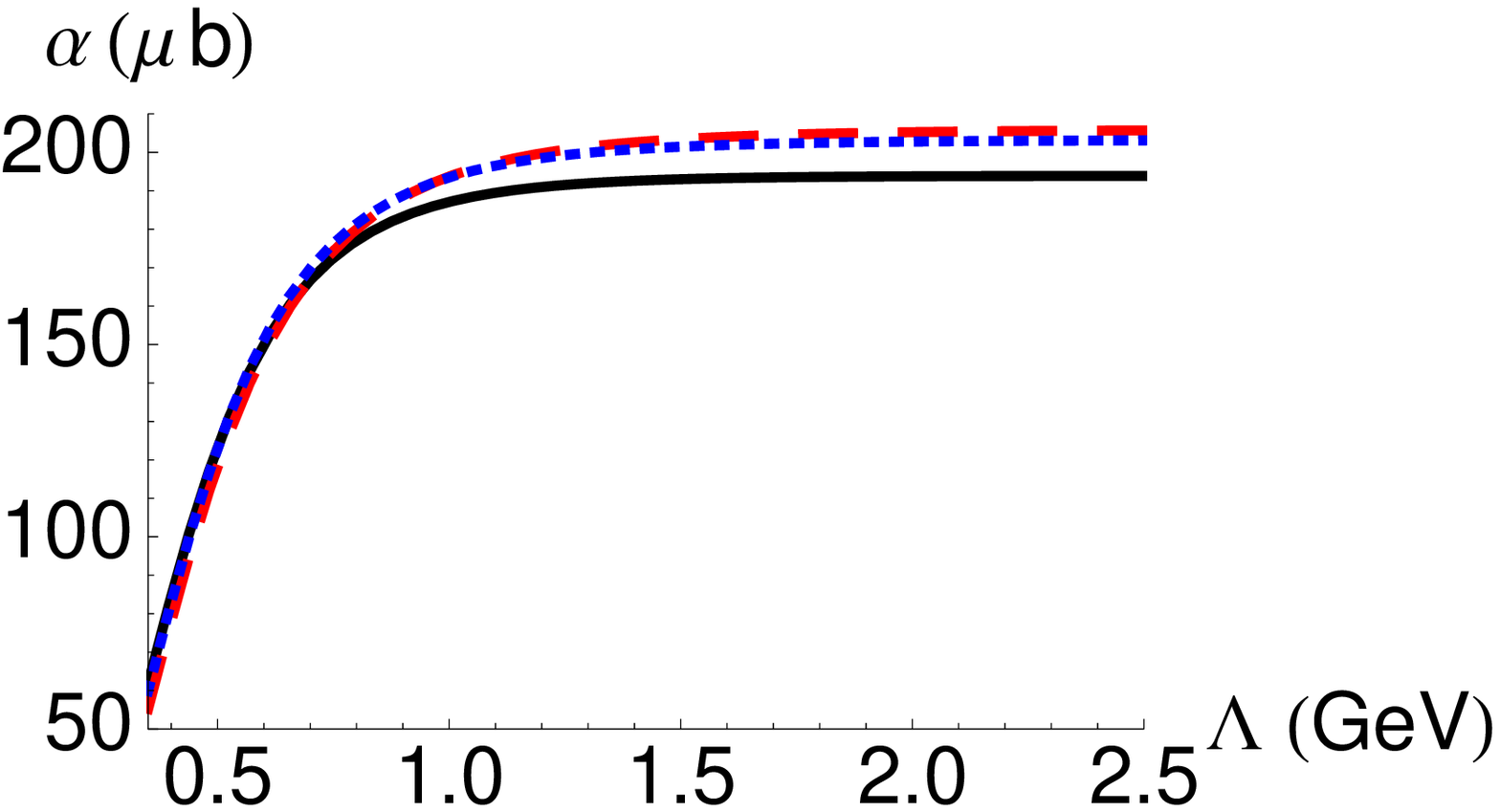}}
\hspace{.1\linewidth}
\subfigure[\ Exact nucleon propagators]{\label{fig:totL-b}\includegraphics[width=.4\linewidth]{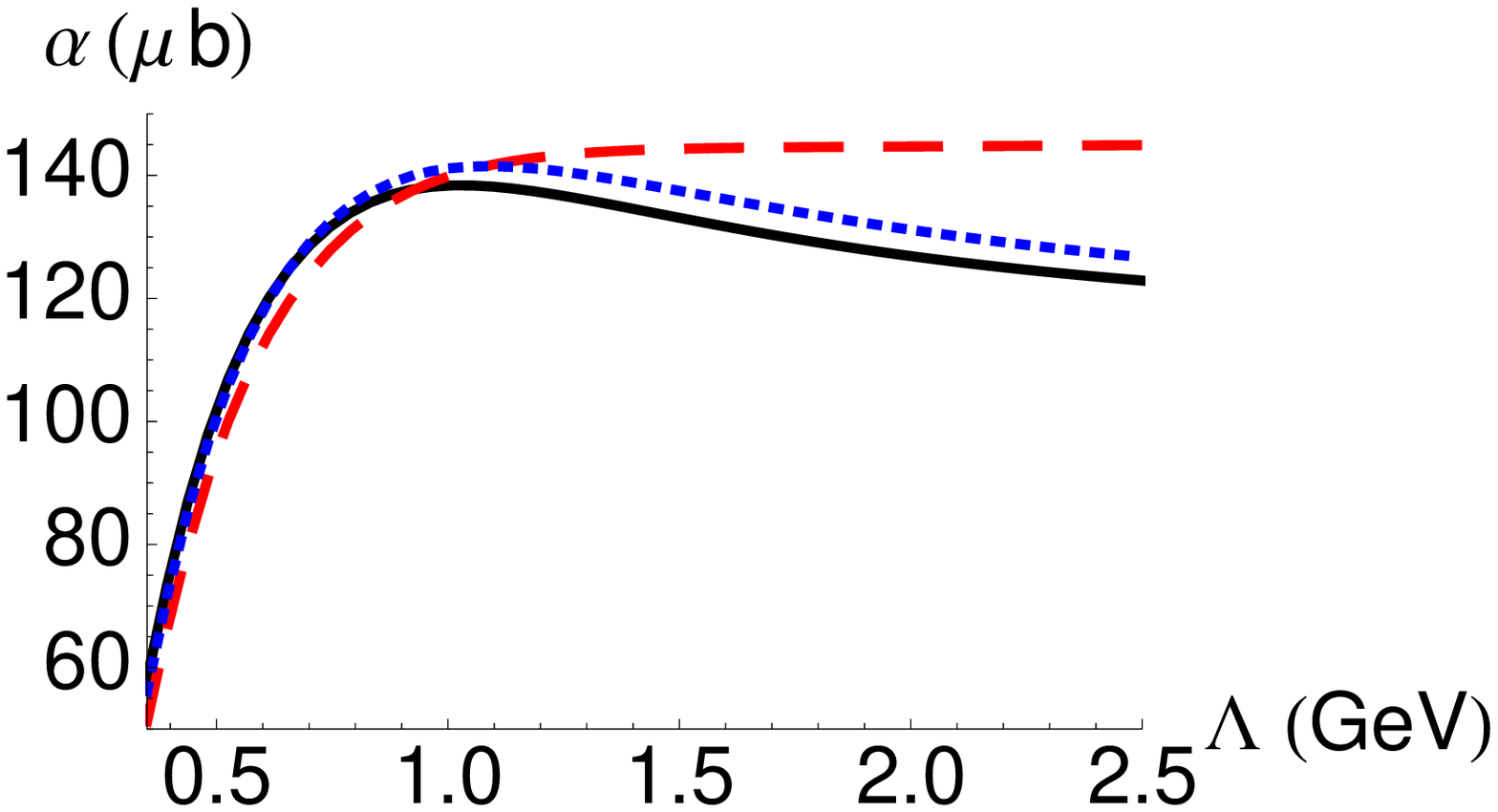}}
\end{center}
\caption{\label{fig:totL}(Color online) Cutoff dependence of the total cross section.  Av18 (solid curve), NijmII (dashed curve), and Reid '93 (dotted curve).}
\label{fig:edge}
\end{figure}

The theoretical total cross section as a function of the cutoff is shown in Fig. \ref{fig:totL}.
The theoretical results include all diagrams up to $\mathcal{O}(\chi^2)$.  Thus theory can assign a rough uncertainty to the threshold cross section of $2\times\chi^2\approx30\%$.

\section{\label{sec:summary}Summary}

Before the findings of this work, the total theoretical cross section at $\Lambda\approx1$ GeV [Fig. \ref{fig:totL-a}] was in agreement with the most recent experiment (fifth row of Table \ref{tab:expt}) at approximately the $1\sigma$ level.  Fixing the FRA approximation decreases the cross section, and if we stop here [Fig. \ref{fig:totL-b}], the agreement between theory and experiment becomes more tenuous (approximately the $2.5\sigma$ level).  A second conclusion regards the MCS power counting, which dictates that $|\vec{p}|\sim\sqrt{m_\pi m_N}$ while $|\vec{k}|\sim m_\pi$.  We find that the wave function corrections of the initial state are of similar size to those of the final state once the FRA is removed, contradicting the previous sentence.  There exists a contact term (an $NNNN\pi$ vertex) at N$^2$LO along with tree-level diagrams proportional to the $c_i$ low-energy constants (LECs) and two-pion exchange loops.  Since all the LECs except the contact term are fixed by other data, it will be interesting to see if that contact term is of natural size.

We plan to investigate further into the concept of reducibility.  Specifically, we would like to define a clear procedure for deciding what to include in the impulse diagram.  There has been a lack of consensus in the literature as to the inclusion of OPE, and if it is included, how that should be done.  Understanding this issue is important, not only for calculation of the total cross section but also for $p$-wave pion production, since the leading contribution to $p$-wave pion production comes from the impulse diagram.

\begin{acknowledgments}
We thank D. Phillips, C. Hanhart, and V. Baru for useful discussions.  This research was supported in part by the U.S. Department of Energy.
\end{acknowledgments}

\begin{appendix}

\section{Lagrange densities\label{sec:lagrangian}}

We define the index of a Lagrange density to be
\be
\nu=d+\frac{f}{2}-2\label{eq:index},
\ee
where $d$ is the sum of the number of derivatives and powers of $m_\pi$, and $f$ is the number of fermion fields.  This represents the standard power counting for nuclear physics.  The $\nu=0$ Lagrangian (with isovectors in $\mathbf{bold}$ font) is \cite{Cohen:1995cc}
\be
\mathcal{L}^{\left(0\right)}=\frac{1}{2}\left(\partial\bpi\right)^2-\frac{1}{2}m_\pi^2\bpi^2+N^\dagger i\partial_0N+\frac{g_A}{2f_\pi}N^\dagger\left(\btau\cdot\vec{\sigma}\cdot\vec{\nabla}\bpi\right)N+...,\label{eq:l0}
\ee
where $\btau$ and $\vec{\sigma}$ are the pauli matrices acting on the isospin and spin of a single nucleon.  The \lqt$+...$" indicates that only the terms used in this calculation are shown.

The $\nu=1$ Lagrangian includes recoil corrections and other terms invariant under $SU(2)_L\times SU(2)_R$.
\be
\mathcal{L}^{\left(1\right)}=\frac{1}{2m_N}N^\dagger\nabla^2N-\frac{1}{2m_N}\left[\frac{g_A}{2f_\pi}iN^\dagger\btau\cdot\dot{\bpi}\vec{\sigma}\cdot\vec{\nabla}N+h.c.\right]+...,\label{eq:l1}
\ee
where we use the values given in Table \ref{tab:parameters}.
\begin{table}
\caption{Parameters used.\label{tab:parameters}}
\begin{center}
\renewcommand{\tabcolsep}{3mm}
\begin{tabular}{|c|c|}
\hline
$m_\pi=134.98\ \text{MeV}$ & $g_A=1.32\ \text{MeV}$\\ \hline
$m_N=938.92\ \text{MeV}$ & $f_\pi=92.4\ \text{MeV}$\\ \hline
\end{tabular}
\end{center}
\end{table}
Note that the terms with the $c_i$ low-energy constants that appear at this order do not get promoted in MCS for these kinematics and are thus not used.  Also, the terms with the $d_i$ low-energy constants do not contribute to $s$-wave pion production.  Finally, the $NNNN$ contact terms, $C_{S,T}$, do not contribute because we are using a potential with a repulsive core [$R_i(r)R_f(r)\rightarrow0$ as $r\rightarrow0$ for $l_i=1$, $l_f=0$].

\section{Impulse approximation details\label{sec:impdetails}}

Evaluating the isospin matrix element
\be
\la00\mid\tau_{1,3}\mid10\ra=1\label{eq:isospin}
\ee
and using the vertex rule shown in Fig. \ref{fig:rules}, we obtain for Fig. \ref{fig:ia}(a) at threshold
\be
\left\la00\mid\hat{\mathcal{M}}'_L(\vec{p},\vec{k})\mid10\right\ra=\frac{g_A}{2f_\pi}\frac{m_\pi}{2m_N}\vec{\sigma}_1\cdot(\vec{p}+\vec{k})(2\pi)^3\delta^3(\vec{p}-\vec{k}),\label{eq:mmom}
\ee
where $\hat{\mathcal{M}}'=\hat{\mathcal{M}}/\sqrt{2m_N\,2m_P\,2m_d}\equiv\hat{\mathcal{M}}/N$.  Since we are using position space $np$ wavefunctions, we Fourier transform the matrix element with respect to $\vec{l}=\vec{k}-\vec{p}$, which is identical to $\vec{q}\,'$ at threshold,
\be
\int\frac{d^3l}{(2\pi)^3}\,e^{i\vec{l}\cdot\vec{r}}(2\pi)^3\delta^3(\vec{p}-\vec{k})=1\label{eq:ft1}
\ee
Note that we group the $\vec{p}$ and $\vec{k}$ with their respective wave functions prior to performing the Fourier transform
\be
\vec{\sigma}_1\cdot(\vec{p}+\vec{k})\rightarrow\vec{\sigma}_1\cdot(-i\overrightarrow{\nabla}_{np}+i\overleftarrow{\nabla}_d).\label{eq:ft2}
\ee
Thus the full position space operator is
\be
\left\la00\mid\hat{\mathcal{M}}'_{L}(\vec{r})\mid10\right\ra=-i\frac{g_A}{2f_\pi}\frac{m_\pi}{2m_N}\vec{\sigma}_1\cdot(\overrightarrow{\nabla}_{np}-\overleftarrow{\nabla}_d).\label{eq:mpos}
\ee
To calculate the diagram with rescattering on the other nucleon, we consider how each part of the left side of Eq. (\ref{eq:mpos}) transforms under $1\leftrightarrow2$.  Since the strong part of the Lagrangian is invariant under isospin, $\hat{\mathcal{M}}$ is invariant.  The initial isospin ket $\left|1,0\right\ra$ is invariant as well, but $\left|0,0\right\ra\rightarrow-\left|0,0\right\ra$ and $\vec{\nabla}\rightarrow-\vec{\nabla}$.  Thus,
\be
\left\la00\mid\hat{\mathcal{M}}'_{L+R}(\vec{r})\mid10\right\ra=-2i\frac{g_A}{2f_\pi}\frac{m_\pi}{2m_N}\vec{S}\cdot(\overrightarrow{\nabla}_{np}-\overleftarrow{\nabla}_d).\label{eq:mpos2}
\ee

The final spin-angle wave function is that of the deuteron, while the initial state for $s$-wave pion production is solely $^3P_1$,
\bea
\mid f(\vec{r})\ra&\equiv&\frac{u(r)}{r}\mid^3S_1\ra+\frac{w(r)}{r}\mid^3D_1\ra\nonumber
\\
\mid i(\vec{r})\ra&\equiv&4\pi i\,\frac{u_{1,1}(r)}{pr}\mid^3P_1\ra,\label{eq:states}
\eea
where we have absorbed the unobservable (since there is only one initial channel available) phase into the definition of the matrix element.  The spin-angle matrix elements are calculated,
\be
\la f(\vec{r})\mid\mid\vec{S}\cdot\left(\overrightarrow{\nabla}-\overleftarrow{\nabla}\right)\mid\mid i(\vec{r})\ra=4\pi i\left[R_f(r)\frac{\partial R_i(r)}{\partial r}+R_{f,2}(r)\frac{2}{r}R_i(r)-\frac{\partial R_f(r)}{\partial r}R_i(r)\right],\label{eq:spinanglegrad}
\ee
where $R_f(r)\equiv\sqrt{2}u(r)/r+w(r)/r$, $R_{f,2}(r)\equiv\sqrt{2}u(r)/r-2w(r)/r$, and $R_i(r)\equiv u_{1,1}(r)/pr$.

Using Eqs. (\ref{eq:mpos2}) and (\ref{eq:spinanglegrad}), we have the final result for the reduced matrix element,
\bea
A_0^\text{imp}&\equiv&\int dr\,r^2\left(\la00\mid\otimes\la f(\vec{r})\mid\mid\right)\hat{\mathcal{M}}(\vec{r})\left(\mid\mid i(\vec{r})\ra\otimes\mid10\ra\right)\nonumber
\\
&=&N8\pi\frac{g_A}{2f_\pi}\frac{m_\pi}{2m_N}K,
\\
K&\equiv&\int dr\,r^2\left[R_f(r)\frac{\partial R_i(r)}{\partial r}+R_{f,2}(r)\frac{2}{r}R_i(r)-\frac{\partial R_f(r)}{\partial r}R_i(r)\right].
\eea

\section{Including OPE details\label{sec:opedetails}}
\subsection{Reducible OPE\label{sec:opedetails1}}

Taking just the $\vec{q}\,'$ terms at the OPE vertices, Fig. \ref{fig:ia}(b) is given by
\bea
\hat{\mathcal{M}}'(\vec{p},\vec{k})&=&\left(-\frac{g_A}{2f_\pi}\right)^3\btau_1\cdot\btau_2\,\vec{\sigma}_1\cdot(-\vec{q}\,')\frac{-i}{\vec{q}\,'\,^2+\mu(0)^2}\vec{\sigma}_2\cdot\vec{q}\,'\nonumber
\\
&&\qquad\qquad\qquad\times\tau_{1,3}\,\frac{i}{-E_d-\vec{p}\,^2/m_N}\vec{\sigma}_1\cdot\left(-\frac{m_\pi}{2m_N}2\vec{p}\right),\label{eq:mwfn}
\eea
where $\mo^2\equiv m_\pi^2-\omega^2$.  Adding to this expression emission from the right nucleon and approximating $\vec{p}\,^2=m_\pi m_N$ as discussed at the end of Sec. \ref{sec:ope}, we find
\bea
\la00\mid\hat{\mathcal{M}}'(\vec{p},\vec{k})\mid10\ra&=&\frac{12g_A^3}{8f_\pi^3}\frac{m_\pi}{2m_N}\vec{\sigma}_1\cdot\vec{q}\,'\vec{\sigma}_2\cdot\vec{q}\,'\frac{1}{\vec{q}\,'\,^2+\mu(0)^2}\frac{1}{-E_d-m_\pi}\vec{S}\cdot\vec{p},\nonumber
\\
\la00\mid\hat{\mathcal{M}}'(\vec{r})\mid10\ra&=&\frac{ig_A^3}{8\pi f_\pi^3}\frac{m_\pi}{2m_N}\mu(0)^3\left(S_{12}f(\omega,r)+\vec{\sigma}_1\cdot\vec{\sigma}_2\,g(\omega,r)\right)\frac{1}{-E_d-m_\pi}\vec{S}\cdot\vec{\nabla},\label{eq:mwfntot}
\eea
where the $\vec{\nabla}$ acts on the initial $np$ wave function, $S_{12}=3\vec{\sigma}_1\cdot\hat{r}\vec{\sigma}_2\cdot\hat{r}-\vec{\sigma}_1\cdot\vec{\sigma}_2$ is the normal tensor operator, and
\bea
g(\omega,r)&=&\frac{e^{-\mo r}}{\mo r},\nonumber
\\
f(\omega,r)&=&\left(1+\frac{3}{\mo r}+\frac{3}{(\mo r)^2}\right)\frac{e^{-\mo r}}{\mo r}
\eea
come from the Fourier transform [see Eq. (\ref{eq:ft1})] of the pion propagator.  Next, we evaluate
\be
\left(\frac{u(r)}{r}\la^3S_1\mid+\frac{w(r)}{r}\la^3D_1\mid\right)\left(S_{12}f(\omega,r)+\vec{\sigma}_1\cdot\vec{\sigma}_2\,g(\omega,r)\right)\equiv\frac{\tilde{u}(\omega,r)}{r}\la^3S_1\mid+\frac{\tilde{w}(\omega,r)}{r}\la^3D_1\mid,\label{eq:ope}
\ee
where
\bea
\frac{\tilde{u}(\omega,r)}{r}=\frac{u(r)}{r}g(\omega,r)+2\sqrt{2}\frac{w(r)}{r}f(\omega,r),\nonumber
\\
\frac{\tilde{w}(\omega,r)}{r}=\frac{w(r)}{r}\left(g(\omega,r)-2f(\omega,r)\right)+2\sqrt{2}\frac{u(r)}{r}f(\omega,r).
\eea
Thus,
\bea
\la f(\vec{r})\mid\mid\left(S_{12}f(\omega,r)+\vec{\sigma}_1\cdot\vec{\sigma}_2\,g(\omega,r)\right)\vec{S}\cdot\vec{\nabla}\mid\mid i(\vec{r})\ra&=&4\pi i\left(\frac{\tilde{u}(\omega,r)}{r}\sqrt{2}\left(\ddr+\frac{2}{r}\right)\right.\nonumber
\\
&&\qquad\left.+\frac{\tilde{w}(\omega,r)}{r}\left(\ddr-\frac{1}{r}\right)\right)R_i(r),\label{eq:redwfn}
\eea
and we finally arrive at the full reduced matrix element,
\bea
A_0^\text{OPE,red,f}&=&-N\frac{g_A^3}{2f_\pi^3}\frac{m_\pi}{2m_N}\frac{\mu(0)^3}{-E_d-m_\pi}L^f(0),\nonumber
\\
L^f(\omega)&=&\int dr\,r^2\left[\frac{\tilde{u}(\omega,r)}{r}\sqrt{2}\left(\ddr+\frac{2}{r}\right)+\frac{\tilde{w}(\omega,r)}{r}\left(\ddr-\frac{1}{r}\right)\right]R_i(r).
\eea

\subsection{Irreducible OPE\label{sec:opedetails2}}

Finally, as described in Sec. \ref{sec:ope}, for the irreducible diagram we use $(-m_\pi/2-\vec{p}\,^2/2m_N)^{-1}\approx(-m_\pi)^{-1}$ for the intermediate nucleon propagator and take $\omega=m_\pi/2$.
\be
A_0^\text{OPE,irr,f}=-N\frac{g_A^3}{2f_\pi^3}\frac{m_\pi}{2m_N}\frac{\mu(m_\pi/2)^3}{-m_\pi}L^f(m_\pi/2).
\ee

\subsection{Initial state OPE\label{sec:opedetails3}}

For OPE in the initial state, the isospin matrix element is $\la 00\mid\tau_{1,3}\btau_1\cdot\btau_2\mid10\ra=1$, and because the initial state consists of just one channel, $^3P_1$,
\be
\left(S_{12}f(\omega,r)+\vec{\sigma}_1\cdot\vec{\sigma}_2\,g(\omega,r)\right)\mid^3P_1\ra=\left(2f(\omega,r)+g(\omega,r)\right)\mid^3P_1\ra.\nonumber
\ee
Evaluating the $\vec{S}\cdot\overleftarrow{\nabla}$ reduced matrix elements, we find
\bea
A_0^\text{OPE,red,i}&=&-N\frac{g_A^3}{2f_\pi^3}\frac{m_\pi}{2m_N}\frac{\mu(0)^3/3}{m_\pi}L^i(0),\nonumber
\\
A_0^\text{OPE,irr,i}&=&-N\frac{g_A^3}{2f_\pi^3}\frac{m_\pi}{2m_N}\frac{\mu(m_\pi/2)^3/3}{m_\pi}L^i(m_\pi/2),\nonumber
\\
L^i(\omega)&=&\int dr\,r^2\left(\sqrt{2}\ddr\frac{u(r)}{r}+\left(\ddr+\frac{3}{r}\right)\frac{w(r)}{r}\right)\left(2f(\omega,r)+g(\omega,r)\right)R_i(r).\label{eq:a0wfni}
\eea

\section{Exact wave function corrections details\label{sec:recoildetails}}

Consider the nucleon propagator for reducible OPE in the initial state.  Pulling out a $-m_N$ and expanding this function in spherical coordinates, we have
\bea
iG_0(\vec{r},\vec{r}\,')&=&-m_N\int\frac{d^3k}{(2\pi)^3}e^{-i\vec{k}\cdot(\vec{r}-\vec{r}\,')}\frac{i}{\vec{k}\,^2-\xi^2-i\epsilon}\nonumber
\\
&=&-2im_N\frac{e^{i\xi |\vec{r}-\vec{r}\,'|}}{4\pi|\vec{r}-\vec{r}\,'|}\nonumber
\\
&=&2m_N\xi\sum_{l,m}j_l(\xi r_<)h_l^{(1)}(\xi r_>)Y^{l\,*}_m(\hat{r}')Y^l_m(\hat{r}),\label{eq:ghelm}
\eea
where $\xi=\sqrt{m_\pi m_N}$ and $r_<(r_>)$ is the lesser (greater) of $|\vec{r}|,|\vec{r}\,'|$.  This spherical partial wave expansion was derived from the differential equation
\be
\left(-\frac{1}{r}\frac{\partial^2}{\partial r^2}r+\frac{l(l+1)}{r^2}-\xi^2\right)\mathcal{G}(r,r')=\frac{\delta(r-r')}{rr'},
\ee
where $iG_0=-im_N\mathcal{G}$.  First, one solves the homogenous equation and requires both finiteness at the origin and outgoing wave behavior for large $r$.  Thus, $\mathcal{G}(r,r')=Aj_l(\xi r_<)h_l^{(1)}(\xi r_>)$.  Next, the boundary condition at $r=r'$ is obtained by integrating the differential equation across the boundary.  In terms of $g(r,r')=rr'\mathcal{G}(r,r')$,
\be
\frac{\partial}{\partial r}g_>(r,r')|_{r=r'+\epsilon}-\frac{\partial}{\partial r}g_<(r,r')|_{r=r'-\epsilon}=-1,
\ee
which yields $A=i\xi$.  At this point in the diagram, the two-nucleon state is still $^3P_1$, so we preform one of the angular integrals and obtain
\be
iG_0(\vec{r},\vec{r}\,')\rightarrow m_N\xi j_1(\xi r_<)h_1^{(1)}(\xi r_>).\label{eq:gfn}
\ee
Thus,
\bea
A_0^\text{OPE,red,i}(m_\pi/2)&=&-N\frac{g_A^3}{2f_\pi^3}\frac{m_\pi}{2m_N}\frac{\mu(0)^3}{3}\left(-im_N\xi L^i(0)\right)\nonumber
\\
L^i(m_\pi/2)&=&\int dr\,dr'\,r^2r'^{2}\left(\sqrt{2}\ddr\frac{u(r)}{r}+\left(\ddr+\frac{3}{r}\right)\frac{w(r)}{r}\right)\nonumber
\\
&&\qquad\qquad\qquad\times j_1(\xi r_<)h_1^{(1)}(\xi r_>)\left(2f(m_\pi/2,r')+g(m_\pi/2,r')\right)R_i(r').
\eea

For the irreducible initial state OPE, the only difference is that a $-2m_N$ gets pulled out and the momentum becomes $\xi'=\sqrt{2m_\pi m_N}$,
\be
A_0^\text{OPE,irr,i}(m_\pi/2)=-N\frac{g_A^3}{2f_\pi^3}\frac{m_\pi}{2m_N}\frac{\mu(m_\pi/2)^3}{3}\left(-2im_N\xi' L^i(m_\pi/2)\right)
\ee
For the final state OPE, we can obtain the correct Green function from Eq. (\ref{eq:gfn}) by letting $\xi\rightarrow i\xi$ and using the correct $l$ for the term under consideration.

\section{Cutoff details\label{sec:cutoffdetails}}

In this section we display the exact expressions needed to implement the Gaussian cutoff of Sec \ref{sec:cutoff}.  For the OPE diagrams, the integral of Eq. (\ref{eq:gLdef}) is evaluated,
\bea
g_\Lambda(\omega,r)&=&\frac{1}{2}e^{\mo^2/\Lambda^2}\left[\frac{e^{-\mo r}}{\mo r}\text{erfc}\left(-\frac{\Lambda r}{2}+\frac{\mo}{\Lambda}\right)\right.\nonumber
\\
&&\qquad\qquad\qquad\left.-\frac{e^{\mo r}}{\mo r}\text{erfc}\left(\frac{\Lambda r}{2}+\frac{\mo}{\Lambda}\right)\right]\label{eq:gcutoff}.
\eea
One also needs derivatives of Eq. (\ref{eq:gcutoff}),
\bea
\vec{\sigma}_1\cdot\vec{\nabla}\vec{\sigma}_2\cdot\vec{\nabla}g_\Lambda(\omega,r)&=&\frac{\mo^3}{3}\left(S_{12}f_\Lambda(\omega,r)+\vec{\sigma}_1\cdot\vec{\sigma}_2l_\Lambda(\omega,r)\right)\nonumber
\\
f_\Lambda(\omega,r)&=&\frac{1}{2}e^{\mo^2/\Lambda^2}\left[\left(1+\frac{3}{\mo r}+\frac{3}{(\mo r)^2}\right)\text{erfc}\left(-\frac{\Lambda r}{2}+\frac{\mo}{\Lambda}\right)\right.\nonumber
\\
&&\left.-\frac{\Lambda}{\sqrt{\pi}\mo}\left(\frac{\Lambda^2}{2\mo^2}\mo r+1+\frac{3}{\mo r}\right)e^{-\left(-\frac{\Lambda r}{2}+\frac{\mo}{\Lambda}\right)^2}\right]\frac{e^{-\mo r}}{\mo r}\nonumber
\\[.1in]
&&+\left(\mu\rightarrow\mu\text{ and }\Lambda\rightarrow-\Lambda\right)\nonumber
\\
l_\Lambda(\omega,r)&=&\frac{1}{2}e^{\mo^2/\Lambda^2}\left[\text{erfc}\left(-\frac{\Lambda r}{2}+\frac{\mo}{\Lambda}\right)\right.\nonumber
\\
&&\left.-\frac{\Lambda}{\sqrt{\pi\mo}}\left(\frac{\Lambda^2}{2\mo^2}\mo r+1\right)e^{-\left(-\frac{\Lambda r}{2}+\frac{\mo}{\Lambda}\right)^2}\right]\frac{e^{-\mo r}}{\mo r}\nonumber
\\[.1in]
&&+\left(\mu\rightarrow\mu\text{ and }\Lambda\rightarrow-\Lambda\right).
\eea

\end{appendix}

\bibliography{references}

\begin{thebibliography}{23}
\expandafter\ifx\csname natexlab\endcsname\relax\def\natexlab#1{#1}\fi
\expandafter\ifx\csname bibnamefont\endcsname\relax
  \def\bibnamefont#1{#1}\fi
\expandafter\ifx\csname bibfnamefont\endcsname\relax
  \def\bibfnamefont#1{#1}\fi
\expandafter\ifx\csname citenamefont\endcsname\relax
  \def\citenamefont#1{#1}\fi
\expandafter\ifx\csname url\endcsname\relax
  \def\url#1{\texttt{#1}}\fi
\expandafter\ifx\csname urlprefix\endcsname\relax\def\urlprefix{URL }\fi
\providecommand{\bibinfo}[2]{#2}
\providecommand{\eprint}[2][]{\url{#2}}

\bibitem[{\citenamefont{Measday and Miller}(1979)}]{Measday:1979if}
\bibinfo{author}{\bibfnamefont{D.~F.} \bibnamefont{Measday}} \bibnamefont{and}
  \bibinfo{author}{\bibfnamefont{G.~A.} \bibnamefont{Miller}},
  \bibinfo{journal}{Ann. Rev. Nucl. Part. Sci.} \textbf{\bibinfo{volume}{29}},
  \bibinfo{pages}{121} (\bibinfo{year}{1979}).

\bibitem[{\citenamefont{Hanhart}(2004)}]{Hanhart:2003pg}
\bibinfo{author}{\bibfnamefont{C.}~\bibnamefont{Hanhart}},
  \bibinfo{journal}{Phys. Rept.} \textbf{\bibinfo{volume}{397}},
  \bibinfo{pages}{155} (\bibinfo{year}{2004}), \eprint{hep-ph/0311341}.

\bibitem[{\citenamefont{Jenkins and Manohar}(1991)}]{Jenkins:1990jv}
\bibinfo{author}{\bibfnamefont{E.~E.} \bibnamefont{Jenkins}} \bibnamefont{and}
  \bibinfo{author}{\bibfnamefont{A.~V.} \bibnamefont{Manohar}},
  \bibinfo{journal}{Phys. Lett.} \textbf{\bibinfo{volume}{B255}},
  \bibinfo{pages}{558} (\bibinfo{year}{1991}).

\bibitem[{\citenamefont{Weinberg}(1991)}]{Weinberg:1991um}
\bibinfo{author}{\bibfnamefont{S.}~\bibnamefont{Weinberg}},
  \bibinfo{journal}{Nucl. Phys.} \textbf{\bibinfo{volume}{B363}},
  \bibinfo{pages}{3} (\bibinfo{year}{1991}).

\bibitem[{\citenamefont{Bernard et~al.}(1993)\citenamefont{Bernard, Kaiser, and
  Meissner}}]{Bernard:1993nj}
\bibinfo{author}{\bibfnamefont{V.}~\bibnamefont{Bernard}},
  \bibinfo{author}{\bibfnamefont{N.}~\bibnamefont{Kaiser}}, \bibnamefont{and}
  \bibinfo{author}{\bibfnamefont{U.~G.} \bibnamefont{Meissner}},
  \bibinfo{journal}{Z. Phys.} \textbf{\bibinfo{volume}{C60}},
  \bibinfo{pages}{111} (\bibinfo{year}{1993}), \eprint{hep-ph/9303311}.

\bibitem[{\citenamefont{Hemmert et~al.}(1998)\citenamefont{Hemmert, Holstein,
  and Kambor}}]{Hemmert:1997ye}
\bibinfo{author}{\bibfnamefont{T.~R.} \bibnamefont{Hemmert}},
  \bibinfo{author}{\bibfnamefont{B.~R.} \bibnamefont{Holstein}},
  \bibnamefont{and} \bibinfo{author}{\bibfnamefont{J.}~\bibnamefont{Kambor}},
  \bibinfo{journal}{J. Phys.} \textbf{\bibinfo{volume}{G24}},
  \bibinfo{pages}{1831} (\bibinfo{year}{1998}), \eprint{hep-ph/9712496}.

\bibitem[{\citenamefont{Cohen et~al.}(1996)\citenamefont{Cohen, Friar, Miller,
  and van Kolck}}]{Cohen:1995cc}
\bibinfo{author}{\bibfnamefont{T.~D.} \bibnamefont{Cohen}},
  \bibinfo{author}{\bibfnamefont{J.~L.} \bibnamefont{Friar}},
  \bibinfo{author}{\bibfnamefont{G.~A.} \bibnamefont{Miller}},
  \bibnamefont{and} \bibinfo{author}{\bibfnamefont{U.}~\bibnamefont{van
  Kolck}}, \bibinfo{journal}{Phys. Rev.} \textbf{\bibinfo{volume}{C53}},
  \bibinfo{pages}{2661} (\bibinfo{year}{1996}), \eprint{nucl-th/9512036}.

\bibitem[{\citenamefont{Lensky et~al.}(2006)}]{Lensky:2005jc}
\bibinfo{author}{\bibfnamefont{V.}~\bibnamefont{Lensky}} \bibnamefont{et~al.},
  \bibinfo{journal}{Eur. Phys. J.} \textbf{\bibinfo{volume}{A27}},
  \bibinfo{pages}{37} (\bibinfo{year}{2006}), \eprint{nucl-th/0511054}.

\bibitem[{\citenamefont{Bolton and Miller}(2010)}]{Bolton:2009rq}
\bibinfo{author}{\bibfnamefont{D.~R.} \bibnamefont{Bolton}} \bibnamefont{and}
  \bibinfo{author}{\bibfnamefont{G.~A.} \bibnamefont{Miller}},
  \bibinfo{journal}{Phys. Rev.} \textbf{\bibinfo{volume}{C81}},
  \bibinfo{pages}{014001} (\bibinfo{year}{2010}), \eprint{0907.0254}.

\bibitem[{\citenamefont{Hanhart et~al.}(2001)\citenamefont{Hanhart, Miller,
  Myhrer, Sato, and van Kolck}}]{Hanhart:2000wf}
\bibinfo{author}{\bibfnamefont{C.}~\bibnamefont{Hanhart}},
  \bibinfo{author}{\bibfnamefont{G.~A.} \bibnamefont{Miller}},
  \bibinfo{author}{\bibfnamefont{F.}~\bibnamefont{Myhrer}},
  \bibinfo{author}{\bibfnamefont{T.}~\bibnamefont{Sato}}, \bibnamefont{and}
  \bibinfo{author}{\bibfnamefont{U.}~\bibnamefont{van Kolck}},
  \bibinfo{journal}{Phys. Rev.} \textbf{\bibinfo{volume}{C63}},
  \bibinfo{pages}{044002} (\bibinfo{year}{2001}), \eprint{nucl-th/0010079}.

\bibitem[{\citenamefont{Park et~al.}(2001)\citenamefont{Park, Kubodera, Min,
  and Rho}}]{Park:2000ct}
\bibinfo{author}{\bibfnamefont{T.-S.} \bibnamefont{Park}},
  \bibinfo{author}{\bibfnamefont{K.}~\bibnamefont{Kubodera}},
  \bibinfo{author}{\bibfnamefont{D.-P.} \bibnamefont{Min}}, \bibnamefont{and}
  \bibinfo{author}{\bibfnamefont{M.}~\bibnamefont{Rho}},
  \bibinfo{journal}{Nucl. Phys.} \textbf{\bibinfo{volume}{A684}},
  \bibinfo{pages}{101} (\bibinfo{year}{2001}), \eprint{nucl-th/0005069}.

\bibitem[{\citenamefont{Wiringa et~al.}(1995)\citenamefont{Wiringa, Stoks, and
  Schiavilla}}]{Wiringa:1994wb}
\bibinfo{author}{\bibfnamefont{R.~B.} \bibnamefont{Wiringa}},
  \bibinfo{author}{\bibfnamefont{V.~G.~J.} \bibnamefont{Stoks}},
  \bibnamefont{and}
  \bibinfo{author}{\bibfnamefont{R.}~\bibnamefont{Schiavilla}},
  \bibinfo{journal}{Phys. Rev.} \textbf{\bibinfo{volume}{C51}},
  \bibinfo{pages}{38} (\bibinfo{year}{1995}), \eprint{nucl-th/9408016}.

\bibitem[{\citenamefont{Stoks et~al.}(1994)\citenamefont{Stoks, Klomp,
  Terheggen, and de~Swart}}]{Stoks:1994wp}
\bibinfo{author}{\bibfnamefont{V.~G.~J.} \bibnamefont{Stoks}},
  \bibinfo{author}{\bibfnamefont{R.~A.~M.} \bibnamefont{Klomp}},
  \bibinfo{author}{\bibfnamefont{C.~P.~F.} \bibnamefont{Terheggen}},
  \bibnamefont{and} \bibinfo{author}{\bibfnamefont{J.~J.}
  \bibnamefont{de~Swart}}, \bibinfo{journal}{Phys. Rev.}
  \textbf{\bibinfo{volume}{C49}}, \bibinfo{pages}{2950} (\bibinfo{year}{1994}),
  \eprint{nucl-th/9406039}.

\bibitem[{\citenamefont{Hanhart and Kaiser}(2002)}]{Hanhart:2002bu}
\bibinfo{author}{\bibfnamefont{C.}~\bibnamefont{Hanhart}} \bibnamefont{and}
  \bibinfo{author}{\bibfnamefont{N.}~\bibnamefont{Kaiser}},
  \bibinfo{journal}{Phys. Rev.} \textbf{\bibinfo{volume}{C66}},
  \bibinfo{pages}{054005} (\bibinfo{year}{2002}), \eprint{nucl-th/0208050}.

\bibitem[{\citenamefont{Ericson and Weise}(1988)}]{Ericson:1988gk}
\bibinfo{author}{\bibfnamefont{T.~E.~O.} \bibnamefont{Ericson}}
  \bibnamefont{and} \bibinfo{author}{\bibfnamefont{W.}~\bibnamefont{Weise}},
  \emph{\bibinfo{title}{Pions and Nuclei}}, vol.~\bibinfo{volume}{74} of
  \emph{\bibinfo{series}{The International Series of Monographs on Physics}}
  (\bibinfo{publisher}{Clarendon}, \bibinfo{address}{Oxford},
  \bibinfo{year}{1988}).

\bibitem[{\citenamefont{Hanhart and Wirzba}(2007)}]{Hanhart:2007mu}
\bibinfo{author}{\bibfnamefont{C.}~\bibnamefont{Hanhart}} \bibnamefont{and}
  \bibinfo{author}{\bibfnamefont{A.}~\bibnamefont{Wirzba}},
  \bibinfo{journal}{Phys. Lett.} \textbf{\bibinfo{volume}{B650}},
  \bibinfo{pages}{354} (\bibinfo{year}{2007}), \eprint{nucl-th/0703012}.

\bibitem[{\citenamefont{Bernard et~al.}(1999)\citenamefont{Bernard, Kaiser, and
  Meissner}}]{Bernard:1998sz}
\bibinfo{author}{\bibfnamefont{V.}~\bibnamefont{Bernard}},
  \bibinfo{author}{\bibfnamefont{N.}~\bibnamefont{Kaiser}}, \bibnamefont{and}
  \bibinfo{author}{\bibfnamefont{U.-G.} \bibnamefont{Meissner}},
  \bibinfo{journal}{Eur. Phys. J.} \textbf{\bibinfo{volume}{A4}},
  \bibinfo{pages}{259} (\bibinfo{year}{1999}), \eprint{nucl-th/9806013}.

\bibitem[{\citenamefont{Park et~al.}(2003)}]{Park:2002yp}
\bibinfo{author}{\bibfnamefont{T.~S.} \bibnamefont{Park}} \bibnamefont{et~al.},
  \bibinfo{journal}{Phys. Rev.} \textbf{\bibinfo{volume}{C67}},
  \bibinfo{pages}{055206} (\bibinfo{year}{2003}), \eprint{nucl-th/0208055}.

\bibitem[{\citenamefont{Gardestig et~al.}(2006)\citenamefont{Gardestig,
  Phillips, and Elster}}]{Gardestig:2005sn}
\bibinfo{author}{\bibfnamefont{A.}~\bibnamefont{Gardestig}},
  \bibinfo{author}{\bibfnamefont{D.~R.} \bibnamefont{Phillips}},
  \bibnamefont{and} \bibinfo{author}{\bibfnamefont{C.}~\bibnamefont{Elster}},
  \bibinfo{journal}{Phys. Rev.} \textbf{\bibinfo{volume}{C73}},
  \bibinfo{pages}{024002} (\bibinfo{year}{2006}), \eprint{nucl-th/0511042}.

\bibitem[{\citenamefont{Hutcheon et~al.}(1990)}]{Hutcheon:1989bt}
\bibinfo{author}{\bibfnamefont{D.~A.} \bibnamefont{Hutcheon}}
  \bibnamefont{et~al.}, \bibinfo{journal}{Phys. Rev. Lett.}
  \textbf{\bibinfo{volume}{64}}, \bibinfo{pages}{176} (\bibinfo{year}{1990}).

\bibitem[{\citenamefont{Heimberg et~al.}(1996)}]{Heimberg:1996be}
\bibinfo{author}{\bibfnamefont{P.}~\bibnamefont{Heimberg}}
  \bibnamefont{et~al.}, \bibinfo{journal}{Phys. Rev. Lett.}
  \textbf{\bibinfo{volume}{77}}, \bibinfo{pages}{1012} (\bibinfo{year}{1996}).

\bibitem[{\citenamefont{Drochner et~al.}(1998)}]{Drochner:1998ja}
\bibinfo{author}{\bibfnamefont{M.}~\bibnamefont{Drochner}} \bibnamefont{et~al.}
  (\bibinfo{collaboration}{GEM}), \bibinfo{journal}{Nucl. Phys.}
  \textbf{\bibinfo{volume}{A643}}, \bibinfo{pages}{55} (\bibinfo{year}{1998}).

\bibitem[{\citenamefont{Strauch et~al.}(2010)}]{Strauch:2010rm}
\bibinfo{author}{\bibfnamefont{T.}~\bibnamefont{Strauch}} \bibnamefont{et~al.},
  \bibinfo{journal}{Phys. Rev. Lett.} \textbf{\bibinfo{volume}{104}},
  \bibinfo{pages}{142503} (\bibinfo{year}{2010}), \eprint{1003.4153}.

\end{thebibliography}

\end{document}